\pdfoutput=1     

\documentclass{IEEEtran}
\usepackage{subfigure}
\usepackage{amssymb}
 \usepackage[pdftex]{graphicx}
\usepackage{url}
\usepackage{color}
\usepackage{setspace}
\usepackage{amsmath}
\usepackage{graphics}
\usepackage{epsfig}
\usepackage{cite}
\newcommand{\defeq}{\stackrel{def}{=}}
\newcommand{\as}{\stackrel{a.s.}{\longrightarrow}}
\newcommand{\bm}[1]{\boldsymbol{#1}}
\newcommand\mb[1]{\mathbf{#1}}
\newcommand{\tr}{\mathrm{tr}}
\newcommand{\var}{\mathrm{Var}}
\newcommand{\MSE}{\mathrm{MSE}}
\newcommand{\SNR}{\mathrm{SNR}}

\newcommand{\Diag}{\mathrm{Diag}}

\newtheorem{theorem}{Theorem}
\newtheorem{lemma}{Lemma}

\newtheorem{remark}{Remark}

\newtheorem{corollary}{Corollary}

\newtheorem{assumption}{Assumption}

\begin{document}

\title{Composite Channel Estimation\\in Massive MIMO Systems}
\author{Ko-Feng Chen, Yen-Cheng Liu, and Yu T. Su*\\
        Institute of Communications Engineering,
        National Chiao Tung University \\1001 Ta-Hsueh Rd.,
        Hsinchu, 30010, Taiwan.\\
        Email: ytsu@mail.nctu.edu.tw
        \thanks{
K.-F. Chen was with the Institute of Communications Engineering,
National Chiao Tung University, Hsinchu, Taiwan and is now with MediaTek
Inc., Hsinchu, Taiwan (email: kevin77923@gmail.com). Y.-C. Liu and Y. T. Su
(correspondence addressee) are with
the Institute of Communications Engineering, National Chiao Tung
University, Hsinchu, Taiwan (email: ycliu@ieee.org; ytsu@nctu.edu.tw).
The material in this paper will be presented in part at the IEEE 2013
GLOBECOM Workshop.}
        }

 \DeclareGraphicsExtensions{.eps}

\maketitle \thispagestyle{empty}

\begin{abstract}
We consider a multiuser (MU) multiple-input multiple-output (MIMO)
time-division duplexing (TDD) system in which the base station
(BS) is equipped with a large number of antennas for communicating
with single-antenna mobile users. In such a system the  BS has to
estimate the channel state information (CSI) that includes large-scale
fading coefficients (LSFCs) and small-scale fading coefficients
(SSFCs) by uplink pilots. Although information about the former
FCs are indispensable in a MU-MIMO or distributed MIMO system,
they are usually ignored or assumed perfectly known when treating
the MIMO CSI estimation problem. We take advantage of the large
spatial samples of a massive MIMO BS to derive accurate LSFC
estimates in the absence of SSFC information. With estimated LSFCs,
SSFCs are then obtained using a rank-reduced (RR) channel model
which in essence transforms the channel vector into a lower dimension
representation.

We analyze the mean squared error (MSE) performance of the
proposed composite channel estimator and prove that the separable
angle of arrival (AoA) information provided by the RR model is
beneficial for enhancing the estimator's performance, especially
when the angle spread of the uplink signal is not too large.
\end{abstract}

\section{Introduction}
A cellular mobile network in which each base station (BS) is
equipped with an $M$-antenna array, is referred to as a
large-scale multiple-input, multiple-output (MIMO) system or a
massive MIMO system for short if $M \gg 1$ and $M \gg K$, where
$K$ is the number of active user antennas within its serving area.
A massive MIMO system has the potentiality of achieving
transmission rate much higher than those offered by current
cellular systems with enhanced reliability and drastically
improved power efficiency. It takes advantage of the so-called
channel-hardening effect \cite{Scaling} which implies that the
channel vectors seen by different users tend to be mutually
orthogonal and frequency-independent \cite{Measurement2}. As a
result, linear receiver is almost optimal in the uplink and simple
multiuser (MU) precoder are sufficient to guarantee satisfactory
downlink performance. Although most investigation consider the
co-located BS antenna array scenario \cite{Scaling}, the use of a
more general setting of massive distributed antennas has been
suggested recently \cite{Lau}.

The Kronecker model \cite{Kron_Model}, which assumes separable
transmit and receive spatial statistics, is often used in the
study of massive MIMO systems \cite{Hoydis}. The spatial channel
model (SCM) \cite{SCM}, which is adopted as the 3GPP standard,
degenerates to the Kronecker model \cite{SCM_v} when the number of
subpaths approaches infinity. This model also implies that the
distributions of angle of arrival (AoA) and angle of departure
(AoD) are independent. In general such an assumption is valid if
the antenna number is small and large cellular system is in
question. But if one side of a MIMO link consists of multiple
single-antenna terminals, only the spatial correlation of the
array side needs to be taken into account and thus the reduced
Kronecker model and other spatial correlated channel models become
equivalent. Throughout this paper our investigation focuses on
this practical scenario, i.e., we consider a massive MIMO system
where $K$ is equal to the number of active mobile users.

We assume that the mobile users transmit orthogonal uplink pilots
for the serving BS to estimate channel state information (CSI)
that includes both small-scale
fading coefficients (SSFCs) and large-scale fading coefficients
(LSFCs). Besides data detection, CSI is needed for a variety of
link adaptation applications such as precoder, modulation and
coding scheme selection. The LSFCs, which summarize the pathloss
and shadowing effect, are proportional to the average received
signal strength (RSS) and are useful in power control, location
estimation, hand-over protocol and other applications. While most
existing works focus on the estimation of the channel matrix which
ignores the LSFC \cite{Yin,Opt_pilot}, it is
desirable to know SSFCs and LSFCs separately. LSFCs are long-term
statistics whose estimation is often more time-consuming than
SSFCs estimation. Conventional MIMO CSI estimators usually assume
perfect LSFC information and deal solely with SSFCs
\cite{Lau,Known_LSFG,Known_LSFG2}. For co-located MIMO systems, it
is reasonable to assume that the corresponding LSFCs remain
constant across all spatial subchannels and the SSFC estimation
can sometime be obtained without the LSFC information. Such an
assumption is no longer valid in a multiuser MIMO (MU-MIMO) system where
the user-BS distances spread over a large range and SSFCs cannot
be derived without the knowledge of LSFCs.

The estimation of LSFC has been largely neglected, assuming
somehow perfectly known prior to SSFC estimation. When one needs
to obtain a joint LSFC and SSFC estimate, the minimum mean square
error (MMSE) or least-squares (LS) criterion is not directly
applicable. The expectation-maximization (EM) approach is a
feasible alternate \cite[Ch. 7]{Kay} but it requires high
computational complexity and cannot guarantee convergence. We
propose an efficient algorithm for estimating LSFCs with no aid of
SSFCs by taking advantage of the channel hardening effect and
large spatial samples available to a massive MIMO BS. Our LSFC
estimator is of low computational complexity, requires relatively
small training overhead, and yields performance far superior to
that of an EM-based estimator. Our analysis shows that it is
unbiased and asymptotically optimal.

Estimation of SSFCs, on the other hand, is more difficult as the
associated spatial correlation is not as high as that among LSFCs.
Nevertheless, given an accurate LSFC estimator, we manage to
derive a reliable SSFC estimator which exploits the spatial
correlation induced channel rank reduction and calls for
estimation of much less channel parameters than that required by
conventional method \cite{Opt_pilot} when the angle spread (AS) of
the uplink signals is small. The proposed SSFC estimator provides
excellent performance and offer additional information about the
average AoA which is very useful in designing a
downlink precoder.

The rest of this paper is organized as follows. In Section
\ref{section:model}, we describe a massive MU-MIMO channel model
that takes into account spatial correlations and large-scale
fading. In Section \ref{section:Estimation}, a novel
uplink-pilot-based LSFC estimator is proposed and in Section
\ref{sec:SSFC}, we devise an SSFC estimator by using the estimated
LSFCs. Simulation results are presented in Section
\ref{section:simulation} to validate the superiority of our rank determination
algorithm and CSI estimators in massive MU-MIMO systems. We
summarize the main contributions in Section
\ref{section:conclusion}.

\textit{Notation:} $(\cdot)^T,(\cdot)^H$, and $(\cdot)^*$
represent the transpose, conjugate transpose, and conjugate of the
enclosed items, respectively. $\mathrm{vec}(\cdot)$ is the
operator that forms one tall vector by stacking columns of the
enclosed matrix, whereas $\Diag(\cdot)$ translates a vector into a
diagonal matrix with the vector entries being the diagonal terms.
While $\mathbb{E}\{\cdot\}$, $\|\cdot\|$, $\|\cdot\|_{2}$, and
$\|\cdot\|_{F}$ denote the expectation, vector $\ell_2$-norm,
matrix spectral norm, and Frobenius norm of the enclosed items,
respectively, $\otimes$ and $\odot$ respectively denote the
Kronecker and Hadamard product operator. Denote by $\mb{I}_L$,
$\mb{1}_L$, and $\mb{0}_L$ respectively the $L\times L$ identity
matrix and $L$-dimensional all-one and all-zero column vectors,
whereas $\mb{1}_{L\times S}$, and $\mb{0}_{L\times S}$ are the
matrix counterparts of the latter two. $\mb{e}_i$ and
$\mb{E}_{ij}$ are all-zero vector and matrix except for their
$i$th and $(i,j)$th element being $1$, respectively.

\section{System Model}
\label{section:model} Consider a single-cell massive MU-MIMO
system having an $M$-antenna BS and $K$
single-antenna mobile stations (MSs), where $M\gg K$. For a
muti-cell system, pilot contamination \cite{Pilot_con} may become
a serious design concern in the worst case when the same pilot
sequences (i.e., the same pilot symbols are placed at the same
time-frequency locations) happen to be used simultaneously in
several neighboring cells and are perfectly synchronized in both
carrier and time. In practice, there are frequency, phase and
timing offsets between any pair of pilot signals and the number of
orthogonal pilots is often sufficient to serve mobile users in
multiple cells. Moreover, neighboring cells may use the same pilot
sequence but the pilot symbols are located in non-overlapping
time-frequency units \cite{LTE}, hence a pilot sequence is more
likely be interfered by uncorrelated asynchronous data sequences
whose impact is not as serious as the worst case and can be
mitigated by proper inter-cell coordination, frequency planning
and some interference suppression techniques \cite{Pilot_con2}. We
will, however, focus on the single-cell narrowband scenario
throughout this paper.

We assume a narrowband communication environment in which a
transmitted signal suffers from both large- and small-scale
fading. The $K$ MS-BS link ranges are denoted by $d_k$ and each
uplink packet place its pilot of length $T$ at the same
time-frequency locations so that, without loss of generality, the
corresponding received samples, arranged in matrix form, ${\bf
Y}=[y_{ij}]$ at the BS can be expressed as
\begin{IEEEeqnarray}{rCl}
 \mb{Y}=\sum_{k=1}^K\sqrt{\beta_k}\mb{h}_k\mb{p}_k^H+\mb{N}
    =\mb{H}\mb{D}_{\boldsymbol{\beta}}^{\frac{1}{2}}\mb{P}+\mb{N}
\label{eq:sys_mod}
\end{IEEEeqnarray}
where $\mathbf{H} = [\mathbf{h}_1, \cdots, \mathbf{h}_K]\in
\mathbb{C}^{M\times K}$ and $\mathbf{D}_{\boldsymbol{\beta}}=
\text{Diag}({\boldsymbol{\beta}})$ contain respectively the SSFCs
and LSFCs that characterize the $K$ uplink channels, and
$\mathbf{N}=[n_{ij}]$ is the white Gaussian noise matrix with
independent identically distributed (i.i.d.) elements, $n_{ij}
\sim \mathcal{CN}(0,1)$. Each element of the vector $\mb{\beta} =
[\beta_1, \cdots, \beta_K]^T$, $\beta_k = {s_k}
d_k^{-\alpha}$, is the product of the random variable
$s_k$ representing the shadowing effect and the path
loss $d_k^\alpha$, $\alpha > 2$. $s_k$ are i.i.d.
log-normal random variables, i.e.,
$10\log_{10}(s_k)\sim \mathcal{N}(0,
\sigma_s^2)$. The $K\times T$ matrix
$\mb{P}=\left[\mb{p}_1,\cdots,\mb{p}_K \right ]^H$, where $T\geq
K$, consists of orthogonal uplink pilot vector $\mb{p}_k$. The
optimality of using orthogonal pilots has been shown in
\cite{Opt_pilot}.

It is reasonable to assume that the mobile users are relatively
far apart (with respect to the wavelength) so that the $k$th
uplink SSFC vector is independent of the $\ell$th vector, $\forall\ell\neq k$, and can be represented by
\begin{IEEEeqnarray}{rCl}
\mb{h}_k=\mb{\Phi}_{k}^{\frac{1}{2}}\tilde{\mb{h}}_k,
\end{IEEEeqnarray}
where $\mb{\Phi}_{k}$ is the spatial correlation matrix at the BS
side with respect to the $k$th user and $\tilde{\mb{h}}_k \sim
\mathcal{CN}(\mb{0}_M,\mb{I}_M)$. We assume that
$\tilde{\mb{h}}_k$'s are i.i.d. and the SSFC $\mb{H}$ remains
constant during a pilot sequence period, i.e., the channel's
coherence time is greater than $T$, while the LSFC $\bm{\beta}$
varies much slower.

\section{Large-Scale Fading Coefficient Estimation}
\label{section:Estimation} Unlike previous works on MIMO channel
matrix estimation which either ignore LSFCs \cite{Yin,Opt_pilot} or assume perfect
known LSFCs \cite{Lau,Known_LSFG,Known_LSFG2}, we try to estimate
$\mb{H}$ and $\mb{D}_{\boldsymbol{\beta}}$ jointly. We first
introduce an efficient LSFC estimator without SSFCs information in
this section. We treat separately channels with and without
spatial correlation at the BS side and show that both cases lead
to same estimators when the BS is equipped with a large-scale
linear antenna array.
\subsection{Uncorrelated BS Antennas}
We first consider the case when the BS antenna spacings are large
enough that the spatial mode correlation is negligible. A
statistic which is a function of the received sample matrix $\mb{Y}$
and LSFCs but is asymptotically independent of the SSFCs is derivable
from the following property \cite[Ch. 3]{RMT}.
\begin{lemma}
Let $\mb{p},\mb{q}\in\mathbb{C}^{M\times 1}$ be two independent
$M$-dimensional complex random vectors with elements i.i.d. as
$\mathcal{CN}(0,1)$. Then by the law of large number
\begin{IEEEeqnarray}{rCl}
\frac{1}{M}\mb{p}^H\mb{p}\stackrel{a.s.}{\longrightarrow}1
~~\text{and}~~
\frac{1}{M}\mb{p}^H\mb{q}\stackrel{a.s.}{\longrightarrow}0
~\text{ as }~M\rightarrow \infty,\notag
\end{IEEEeqnarray}
where $\as$ denotes almost surely convergence.
\label{lemma:LLN}
\end{lemma}
For a massive MIMO system with $M\gg T\geq K$, we have, as
$M\rightarrow\infty$, $\frac{1}{M}\mb{H}^H\mb{H}\as\mb{I}_K$,
$\frac{1}{M} \mb{N}^H\mb{N}\as\mb{I}_T$,
$\frac{1}{M}\mb{H}^H\mb{N}\as \mb{0}_{K\times T}$, and thus
\begin{IEEEeqnarray}{rCl}
    \frac{1}{M}\mb{Y}^H\mb{Y}-\mb{I}_T&=&
    \mb{P}^H\mb{D}_{\boldsymbol{\beta}}\mb{P}
    +\frac{1}{M}\mb{N}^H\mb{N}-\mb{I}_T\notag\\
    &&+\:\mb{P}^H\mb{D}_{\boldsymbol{\beta}}^{\frac{1}{2}}
    \left(\frac{1}{M}\mb{H}^H\mb{H}-\mb{I}_K\right)
    \mb{D}_{\boldsymbol{\beta}}^{\frac{1}{2}}\mb{P}~~~~~\notag\\
    &&+\:\frac{2}{M}\mathfrak{R}\left\{\mb{P}^H\mb{D}_{\boldsymbol{\beta}}^{\frac{1}{2}}
    \mb{H}^H\mb{N}\right\}\notag\\
    &\stackrel{a.s.}{\longrightarrow}&
    \mb{P}^H\mb{D}_{\boldsymbol{\beta}}\mb{P}
    \label{eqn:YHY_appr}
\end{IEEEeqnarray}
(\ref{eqn:YHY_appr}) indicates that the additive noise effect is
reduced and the estimation of LSFCs can be decoupled from that of
the SSFCs. Using the identity, $\mathrm{vec}(\mb{A}\cdot
\Diag(\mb{c}) \cdot \mb{F})=\left((\mb{1}_S\otimes
\mb{A})\odot(\mb{F}^T\otimes \mb{1}_T)\right)\mb{c}$ with
$\mb{A}\in\mathbb{C}^{T\times K}$, $\mb{F}\in\mathbb{C}^{K\times
S}$, and $\mb{c}\in\mathbb{C}^{K\times 1}$, we simplify
(\ref{eqn:YHY_appr}) as
\begin{IEEEeqnarray}{rCl}
    \mathrm{vec}\left(\frac{1}{M}\mb{Y}^H\mb{Y}
    -\mb{I}_T\right)
    \as
    \left(\left(\mb{1}_T\otimes \mb{P}^H\right)\odot
    \left(\mb{P}^T\otimes \mb{1}_T\right)\right)\bm{\beta}.\notag
\end{IEEEeqnarray}
This equation suggests that we solve the following unconstrained
convex problem
\begin{IEEEeqnarray}{rCl}
\underset{\boldsymbol{\beta}}{\min}&&\Bigg\|\mathrm{vec}\left(\frac{1}{M}\mb{Y}^H\mb{Y}
    -\mb{I}_T\right)\notag\\
&&\hspace{7em}
-\left(\left(\mb{1}_T\otimes \mb{P}^H\right)\odot
    \left(\mb{P}^T\otimes \mb{1}_T\right)\right)\boldsymbol{\beta}\Bigg\|^2,\notag
\end{IEEEeqnarray}
to obtain the LSFC estimate
\begin{IEEEeqnarray}{rCl}
\hat{\bm{\beta}}&=&\Diag\left({\|\mb{p}_1\|^{-4}},\cdots,
 {\|\mb{p}_K\|^{-4}}\right) \nonumber\\
    &&~\cdot \left((\mb{1}_T^T\otimes \mb{P})\odot(\mb{P}^*\otimes \mb{1}_T^T)\right)
     \mathrm{vec}\left(\frac{1}{M}\mb{Y}^H\mb{Y}-\mb{I}_T\right).~~~
\label{ALS}
\end{IEEEeqnarray}
This LSFC estimator is of low complexity as no matrix inversion is
needed when orthogonal pilots are used and does not require any
knowledge of SSFCs. Furthermore, the configuration of massive MIMO
makes the estimator robust against noise, which is verified
numerically later in Section \ref{section:simulation}.

\subsection{Correlated BS Antennas}
In practice, the spatial correlations are nonzero and $\mb{Y}$
is of the form
\begin{IEEEeqnarray}{rCl}
\label{eqn:cor_equation}
    \mb{Y}=\tilde{\mb{\Phi}}
    \small{\left[\begin{array}{ccc}
            \tilde{\mb{h}}_{1}  & \cdots & 0\\
            \vdots & \ddots  & \vdots\\
            0 & \cdots  & \tilde{\mb{h}}_{K}
            \end{array}\right]}
    \mb{D}_{\boldsymbol{\beta}}^{\frac{1}{2}}\mb{P}+\mb{N}
\defeq\tilde{\mb{\Phi}}\tilde{\mb{H}}\mb{D}_{\boldsymbol{\beta}}^{\frac{1}{2}}\mb{P}+\mb{N}
\notag
\end{IEEEeqnarray}
where
$\tilde{\mb{\Phi}}=[\mb{\Phi}_{1}^{\frac{1}{2}},\cdots,
\mb{\Phi}_{K}^{\frac{1}{2}}]$. Following \cite{Hoydis,Huh},
we assume that the following is always satisfied:
\begin{assumption}
The spatial correlation at BS antennas seen by a user satisfies
\begin{IEEEeqnarray}{rCl}
\underset{{M\rightarrow\infty}}{\limsup}\|\mb{\Phi}_{k}^{\frac{1}{2}}\|_2<\infty, ~~\forall k; \notag
\end{IEEEeqnarray}
or equivalently,
\begin{IEEEeqnarray}{rCl}
\underset{{M\rightarrow\infty}}{\limsup}\|\mb{\Phi}_{k}\|_2<\infty,
~~\forall k. \notag
\end{IEEEeqnarray}
\label{assump:1}
\end{assumption}
Therefore, (\ref{eqn:YHY_appr}) becomes
\begin{IEEEeqnarray}{rCl}
    \frac{1}{M}\mb{Y}^H\mb{Y}-\mb{I}_T&\as&
    \mb{P}^H\mb{D}_{\boldsymbol{\beta}}\mb{P}
    +\frac{2}{M}\mathfrak{R}\left\{\mb{P}^H\mb{D}_{\boldsymbol{\beta}}^{\frac{1}{2}}
    \tilde{\mb{H}}^H\tilde{\mb{\Phi}}^H\mb{N}\right\}\notag\\
    &&+\:\mb{P}^H\mb{D}_{\boldsymbol{\beta}}^{\frac{1}{2}}
    \left(\frac{1}{M}\tilde{\mb{H}}^H\tilde{\mb{\Phi}}^H
 \tilde{\mb{\Phi}}\tilde{\mb{H}}-\mb{I}_K\right)
    \mb{D}_{\boldsymbol{\beta}}^{\frac{1}{2}}\mb{P}~~~\notag\\
 &\defeq&\mb{P}^H\mb{D}_{\boldsymbol{\beta}}\mb{P}+\mb{N}'\notag
\end{IEEEeqnarray}
where $\mb{N}'$ is zero-mean with seemingly non-diminishing
variance due to the spatial correlation. Nonetheless, we proved in
Appendix A that
\begin{theorem}
If $\underset{{M\rightarrow\infty}}{\limsup}\underset{{1\leq k\leq
K}}{\sup}\|\mb{\Phi}_{k}^{\frac{1}{2}}\|_2<\infty$, then
\begin{IEEEeqnarray}{rCl}
\frac{1}{M}\tilde{\mb{H}}^H\tilde{\mb{\Phi}}^H\tilde{\mb{\Phi}}\tilde{\mb{H}}
&\as&\mb{I}_K\label{H_Phi_as},\\
\frac{1}{M}\tilde{\mb{H}}^H\tilde{\mb{\Phi}}^H\mb{N}
&\as&\mb{0}_{K\times T}\label{H_Phi_N_as}
\end{IEEEeqnarray}
as $M\rightarrow\infty$.
\label{prop:Mean_phi}
\end{theorem}
This theorem implies that although the nonzero spatial correlation does
cause the increase of variance of $\mb{N}'$, the channel hardening
effect still exist and $\mb{N}'$ is asymptotically diminishing provided
that {\it Assumption \ref{assump:1}} holds.
In this case, LS criterion
also mandates the same estimator as (\ref{ALS}).
Several remarks are worth mentioning.
\begin{remark}
\label{note:als} If $J$ consecutive coherence blocks in which the
LSFCs remain constant are available, (\ref{ALS}) can be rewritten
as
\begin{IEEEeqnarray}{rCl}
\hat{\bm{\beta}}&=&\Diag\left({\|\mb{p}_1\|^{-4}},\cdots,
 {\|\mb{p}_K\|^{-4}}\right)
     \left[\left(\mb{1}_T^T\otimes \mb{P}\right)\odot
     \left(\mb{P}^*\otimes \mb{1}_T^T\right)\right]\nonumber\\
    &&~~~~~~~~~~~~~~~~~~~~~\cdot \mathrm{vec}\left(\frac{1}{MJ}
    \sum_{i=1}^J\mb{Y}_i^H\mb{Y}_i-\frac{1}{J}\mb{I}_T\right)
    \label{ALS_J}
\end{IEEEeqnarray}
where $\mb{Y}_i$ is the $i$th received block. Moreover, the noise
reduction effect becomes more evident as more received samples
become available.
\end{remark}
\begin{remark}
\label{note:decouple} The proposed LSFC estimators (\ref{ALS}) and
(\ref{ALS_J}) render element-wise expressions as
\begin{IEEEeqnarray}{rCl}
\hat{\beta}_k&=&\frac{\mb{p}_k^H\mb{Y}^H\mb{Y}\mb{p}_k-M\|\mb{p}_k\|^2}{M\|\mb{p}_k\|^4},
~~\forall ~k,
\label{ALS_decouple}\\
\hat{\beta}_k&=& \frac{\sum_{i=1}^J\mb{p}_k^H\mb{Y}^H_i\mb{Y}_i
\mb{p}_k-MJ\|\mb{p}_k\|^2}{MJ\|\mb{p}_k\|^4}, ~~\forall~ k.
\label{ALS_J_decouple}
\end{IEEEeqnarray}
\end{remark}

\subsection{Performance Analysis}
\label{sec:LSFCanalysis}
Since the mean of the LSFC estimator (\ref{ALS_decouple})
\begin{IEEEeqnarray}{rCl}
\mathbb{E}\left\{\hat{\beta}_k\right\}&=&
\frac{\mb{p}_k^H(M\mb{P}^H\mb{D}_{\boldsymbol{\beta}}\mb{P}+M\mb{I}_K)
\mb{p}_k-M\|\mb{p}_k\|^2}{M\|\mb{p}_k\|^4}\nonumber\\
&=&\frac{M\beta_k\|\mb{p}_k\|^4+M\|\mb{p}_k\|^2-M\|\mb{p}_k\|^2}{M\|\mb{p}_k\|^4}\nonumber\\
&=&\beta_k,~\forall ~k,
\end{IEEEeqnarray}
the mean squared error (MSE) of $\hat{\beta}_k$ 
\begin{IEEEeqnarray}{rCl}
\MSE\left(\hat{\beta}_k\right)=\mathbb{E}\left\{\left|\hat{\beta}_k-\beta_k \right|^2 \right\}
&=&
\var\left\{\hat{\beta}_k\right\}. 
\end{IEEEeqnarray}
\begin{lemma}\cite[Th. 3.4]{RMT}
Let $\mb{A}\in \mathbb{C}^{M\times M}$ and $\mb{p}$ and $\mb{q}$
be two vectors whose elements are i.i.d. as $\mathcal{CN}(0,1)$.
If $\underset{{M\rightarrow\infty}}{\limsup}\|\mb{A}\|_2<\infty$,
then
\begin{IEEEeqnarray}{rCl}
\mb{p}^H\mb{A}\mb{p}\as \tr(\mb{A})
~~\text{and}~~
\frac{1}{M}\mb{p}^H\mb{A}\mb{q}\as 0
~\text{ as }~M\rightarrow \infty.\notag
\end{IEEEeqnarray}
\label{lemma:trace}
\end{lemma}
\begin{remark}
Using \cite[Lemma B.26]{Bai}, we can prove that the convergence
rates in the aforementioned asymptotic formulae follow
$\mathcal{O}({\|\mathbf{A}\|_F}/{M})$. More precisely,
\begin{IEEEeqnarray}{rCl}
\mathbb{E}\left\{\left|\frac{\mb{p}^H\mb{A}\mb{p}-
\tr(\mb{A})}{M}\right|\right\}&=& \mathcal{O}({\|\mathbf{A}\|_F}/{M});\\
\mathbb{E}\left\{\left|\frac{\mb{p}^H\mb{A}\mb{q}}{M}\right|\right\}&=&
\mathcal{O}({\|\mathbf{A}\|_F}/{M}).
\end{IEEEeqnarray}
\label{remark:converge_rate}
\end{remark}
By reformulating (\ref{ALS_decouple}) as
\begin{IEEEeqnarray}{rCl}
\hat{\beta}_k=\beta_k+\underbrace{\frac{\mb{p}_k^H
(\mb{N}^H\mb{N}-M\mb{I}_K)\mb{p}_k}{M\|\mb{p}_k\|^4}}_{r_1}
&&\notag\\
+\underbrace{\frac{\beta_k(\mb{h}_k^H\mb{h}_k-M)}{M}}_{r_2}
&+&\underbrace{\frac{\sqrt{\beta_k}\left(2\mathfrak{R}\left\{
\mb{h}_k^H\mb{N}\mb{p}_k
\right\}\right)}{M\|\mb{p}_k\|^2}}_{r_3},\notag
\end{IEEEeqnarray}
and invoking \textit{Assumption \ref{assump:1}}, \textit{Lemmas
\ref{lemma:LLN}} and \textit{\ref{lemma:trace}}, and the fact that
$\mb{h}_k=\mb{\Phi}_k^{\frac{1}{2}}\tilde{\mb{h}}_k$, we conclude
that $r_1,r_2,r_3\as0$ as $M\rightarrow\infty$, and thus
\begin{IEEEeqnarray}{rCl}
\var\left\{\hat{\beta}_k\right\} 
&=&\mathbb{E}\left\{|r_1+r_2+r_3|^2\right\}\as 0.
\end{IEEEeqnarray}
As $\beta_k$, $\mb{h}_k$, and $\mb{N}$ are uncorrelated, we have
\begin{IEEEeqnarray}{rCl}
\mathbb{E}\left\{|r_1+r_2+r_3|^2\right\}\approx\mathbb{E}
\left\{|r_1|^2\right\}+\mathbb{E}\left\{|r_2|^2\right\}+\mathbb{E}
\left\{|r_3|^2\right\}.\nonumber\\
\label{eqn:17}
\end{IEEEeqnarray}
Since when pilot length $T=K$,
\begin{IEEEeqnarray}{rCl}
\mathbb{E}\left\{|r_1|^2\right\}&=&\|\mb{p}_k\|^{-8}
\mathbb{E}\left\{\tr\left(\mb{p}_k\mb{p}_k^H\left(\frac{\mb{N}^H
\mb{N}-M\mb{I}_K}{M}\right)\right)^2\right\}\nonumber\\
&\leq&\|\mb{p}_k\|^{-8}\mathbb{E}\left\{\tr\left(\mb{p}_k\mb{p}_k^H\right)^2
\tr\left(\frac{\mb{N}^H\mb{N}-M\mb{I}_K}{M}\right)^2\right\}\nonumber\\
&=&K\|\mb{p}_k\|^{-4}
\mathbb{E}\left\{\left(\frac{\mathbf{n}_1^H
\mathbf{n}_1-M}{M}\right)^2\right\}\nonumber\\
&=& \mathcal{O}\left(\frac{K}{\|\mb{p}_k\|^4}\frac{1}{M}\right)
= \mathcal{O}\left(\frac{\beta_k^2}{T\cdot\SNR_k^2}\frac{1}{M}\right);\\
\mathbb{E}\left\{|r_2|^2\right\}&=&\beta_k^2\:\mathbb{E}\left\{\left|\frac{\tilde{\mb{h}}_k^H
\mb{\Phi}_k\tilde{\mb{h}}_k-M}{M}\right|^2\right\}
\nonumber\\&=&
\mathcal{O}\left(\beta_k^2\frac{\|\mb{\Phi}_k\|_F^2}{M^2}\right)
\gg
\mathcal{O}\left(\beta_k^2\frac{1}{M}\right);\\
\mathbb{E}\left\{|r_3|^2\right\}
\hspace{-.4em}&=&\hspace{-.4em}\frac{4\beta_k}{\|\mb{p}_k\|^4}\mathbb{E}
\left\{\left|\frac{\mathfrak{R}\left\{
\mb{h}_k^H\mb{N}\mb{p}_k
\right\}}{M}\right|^2\right\} \nonumber\\
\hspace{-.4em}&\leq&\hspace{-.4em}\frac{4\beta_k}{\|\mb{p}_k\|^4}\mathbb{E}
\left\{\hspace{-.2em}\left|\frac{
\mb{h}_k^H\mb{N}\mb{p}_k}{M}\right|^2\hspace{-.2em}\right\}
\hspace{-.4em}\leq\hspace{-.4em}\frac{4\beta_k}{\|\mb{p}_k\|^2}\mathbb{E}
\left\{\hspace{-.2em}\frac{\left\|
\mb{N}^H\mb{h}_k\right\|^2}{M^2}\hspace{-.2em}\right\}\nonumber\\
\hspace{-.4em}&=&\hspace{-.4em}\frac{4K\beta_k}{\|\mb{p}_k\|^4}
\mathbb{E}\left\{
\left|\frac{\mathbf{n}_1^H\mb{\Phi}_k^{\frac{1}{2}}
\tilde{\mathbf{h}}_k}{M}\right|^2\right\}\nonumber\\
\hspace{-.4em}&=&\hspace{-.4em} \frac{4K\beta_k}{\|\mb{p}_k\|^2}\mathcal{O}\left(\frac{\|\mb{\Phi}_k^\frac{1}{2}\|_F^2}{M^2}\right)
\hspace{-.4em}=\hspace{-.2em}\mathcal{O}\left(\frac{4\beta_k^2}{\SNR_k}\frac{1}{M}\right),
\end{IEEEeqnarray}
where $\SNR_k\defeq {\beta_k\|\mb{p}_k\|^2}/{T}$, we obtain the following
\begin{lemma}
The convergence rate for the MSE of LSFC estimate $\hat{\beta}_k$
is dominated by the term $\mathbb{E}\left\{|r_2|^2\right\}$ when $T=K$.
\label{lemma:convergence_rate}
\end{lemma}

\begin{corollary}
The LSFC estimators (\ref{ALS}) and (\ref{ALS_J}) approach the
minimum mean-square error (MMSE) estimator with asymptotically
diminishing MSE as $M\rightarrow\infty$.
\end{corollary}
\begin{remark}
As $\mathbb{E}\left\{|r_2|^2\right\}$ is the only term
in (\ref{eqn:17}) related to spatial correlation, 
for cases with finite $M$, the MSE-minimizing spatial
correlation matrix $\mb{\Phi}_k^{\star}$ is the solution of
\begin{IEEEeqnarray}{rCl}
\min_{\mb{A}}&&~ ~ \mathbb{E}\left\{\left|\tilde{\mb{h}}_k^H
\mb{A}\tilde{\mb{h}}_k\right|^2\right\}-\tr(\mb{A})\nonumber\\
\mathrm{s.t.}&&~~[\mb{A}]_{ii}=1,~\forall~ i.
\label{prob:17}
\end{IEEEeqnarray}
Following the method of Lagrange multiplier, we obtain $\mb{\Phi}_k^{\star}=\mb{I}_M$. The
convexity of (\ref{prob:17}) implies that $\var\{\hat{\beta}_k\}$
is an increasing function of $\|\mb{\Phi}_k-\mb{I}_M\|_F$, i.e.,
the MSE of the LSFC estimator decreases as the channel
becomes less correlated; meanwhile, \textit{Lemma
\ref{lemma:convergence_rate}} 
guarantees the error convergence rate improvement.
\end{remark}
\begin{remark}[Finite $M$ scenario]
Low MSE, in the order of $10^{-5}$ to $10^{-4}$ if normalized by
LSFCs' variance,
is obtainable with not-so-large BS antenna numbers (e.g., $50$).
The above MSE performance analysis is validated via simulation
in Section \ref{section:simulation}.
\end{remark}

\section{Estimation of Small-Scale Fading Coefficients }
\label{sec:SSFC} Since the SSFC estimation scheme is valid for any
user-BS link, for the sake of brevity, we omit the user index $k$
in the ensuing discussion.
\subsection{Reduced-Rank Channel Modeling}
\label{subsec:SSFC}

In \cite{Model_based}, two analytic correlated MIMO channel models
were proposed. These models generalize and encompass as special
cases, among others, the Kronecker \cite{Kron_Model,sim_model},
virtual representation \cite{Sayeed} and Weichselberger
\cite{Ozcelik} models. They often admit flexible reduced-rank
representations. Moreover, if the AS of the
transmit signal is small, which, as reported in a recent
measurement campaign \cite{Measurement2}, is the case when a large
uniform linear array (ULA) is used at the BS, one of the models
can provide AoA information. In other words,
since the ASs from uplink users in a massive MIMO system are
relatively small (say, less than $15^{\circ}$), the following rank-reduced (RR)
model is easily derivable from \cite{Model_based}
\begin{lemma}[RR representations]
\label{lemma:reducerank}
The channel vector seen by $k$th user can be represented by
\begin{equation}
\mb{h}=\mb{Q}_m^{(\textrm{I})}\mb{c}^{(\textrm{I})}
\label{C_only}
\end{equation}
or alternately by
\begin{equation}
\mb{h}=\mb{W}(\phi)\mb{Q}_m^{(\textrm{II})}\mb{c}^{(\textrm{II})}
\label{RRApp}
\end{equation}
where $\mb{Q}_m^{(\textrm{I})}, \mb{Q}_m^{(\textrm{II})} \in
\mathbb{C}^{M\times m}$ are predetermined basis (unitary) matrices
and $\mb{c}^{(\textrm{I})}, \mb{c}^{(\textrm{II})} \in
\mathbb{C}^{m\times 1}$ are the transformed channel vectors with respect
to bases $\mb{Q}_m^{(\textrm{I})}$ and $\mb{Q}_m^{(\textrm{II})}$
for the user $k$-BS link and $\mb{W}(\phi)$ is diagonal with unit
magnitude entries. The two equalities hold only if $m=M$ and
become approximations if $m < M$.
\end{lemma}
\begin{remark}
It was shown \cite{Model_based} that for an uniform linear array (ULA)
with antenna spacing $\xi$ and incoming signal wavelength $\lambda$,
if $[\mb{W}(\phi)]_{ii}=\exp\left(-j2\pi\frac{(i-1)\xi}{\lambda} \sin\phi\right)$,
then $\phi$ can be interpreted as the mean AoA with respect to the
ULA broadside when AS of ${\bf h}$ is
assumed to be small. A direct implication is that the mean AoA (which
is approximately equal to the incident angle of the strongest path) of
each user link is extractable if the associated AS is small.
(See \ref{app:sp_corr}.)
\label{remark:Adv_II_1}
\end{remark}
\begin{remark}
The measurement reported in \cite{Measurement2} verified that the AS
for each MS-BS link is indeed relatively small when the BS is equipped
with a large-scale linear array. Hence, a massive MIMO channel estimator
based on the model (\ref{RRApp}) is capable of offering accurate mean AoA
information \cite{Scaling} which can then be used by the BS to perform
downlink beamforming.\label{remark:Adv_II_2}
\end{remark}
\begin{remark}
(21) implies that for the user $k$ to BS link, we are interested in
estimating the transformed vector $\mathbf{c}_k={\bf Q}^H_m\mathbf{W}^H(\phi_k){\bf h}_k$
which is obtained by realigning ${\bf h}_k$ and transform it into a new
orthogonal coordinate. The best dimension reduction is obtained by setting ${\bf Q}_m$
as the one that consists of the eigenvectors associated with the largest $m$
eigenvalues of the expected Gram matrix $\mathbb{E}\left\{\mathbf{W}^H(\phi_k)
{\bf h}_k{\bf h}_k^H\mathbf{W}(\phi_k)\right\}=\mathbf{W}^H(\phi_k)\mathbb{E}\{
{\bf h}_k{\bf h}_k^H\}\mathbf{W}(\phi_k)=\mathbf{W}^H(\phi_k)\Phi_k\mathbf{W}(\phi_k)$.
\end{remark}
\begin{remark}
The use of predetermined unitary matrices ${\bf Q}_m$ in both (\ref{C_only})
and (\ref{RRApp}) avoids the estimation of the above correlation matrix
$\mb{\Phi}_k$ and the ensuing eigen-decomposition for each $k$ to obtain the
associated Karhunen-Lo\`{e}ve transform (KLT) basis (eigen vectors).
For large-scale ULAs, due to space limitation, the spatial correlation can be
high and small $m$ is sufficient to capture the spatial variance of the SSFCs if an
appropriate basis matrix is preselected. This is also validated via
simulation in Section \ref{section:simulation}. The advantages of (\ref{RRApp})
with respect to (\ref{C_only}) are that the former can offer additional AoA
information when AS is relatively small and because of the extra alignment operation
$\mathbf{W}^H(\phi_k)$, it makes the resulting Gram matrix closer to a real matrix.
\label{remark:Adv_II_3}
\end{remark}

\subsection{Predetermined Basis for RR Channel Modeling}
\label{subsec:basis}
In addition, KLT basis is nonflexible in that
it is channel-dependent and computationally expensive to obtain.
Thus, prior to the SSFC estimation, eigen-decomposition and
eigenvalue ordering must be
performed to
the spatial correlation matrix of $\mb{W}^H(\phi)\mb{h}$,
which varies from user to user and can be accurately
estimated only if
sufficient observations are collected.
As a result, it is unrealistic to apply
KLT bases in the multiuser SSFC estimation.
Our channel model (\ref{RRApp}) uses a predetermined
signal-independent basis $\mathbf{Q}_m$ which requires far less complexity.
Two candidate bases are of special interest to us for their proximity
to the KLT basis.
\subsubsection{Polynomial Basis \cite{Model_based}}
As the BS antenna spatial correlation is often reasonably smooth,
polynomial basis of dimension $m<M$ may be sufficient to track
the channel variation. To construct an orthonormal discrete polynomial
basis we perform standard QR decomposition
$\mb{U}=\mb{Q}\mb{R}$, where $[\mb{U}]_{ij}=(i-1)^{j-1}$, $\forall
i,j=1,\cdots,M$. Since the polynomial degree of each column of
$\mb{Q}$ are arranged in an ascending order, the RR basis
$\mb{Q}_m$ is obtained by keeping the first $m$ columns.
\subsubsection{Type-$2$ Discrete Cosine Transform (DCT)
Basis \cite{DCT}} DCT, especially Type-$2$ DCT (DCT-$2$ or simply
DCT), is a widely used for image coding for its excellent energy
compaction capability \cite{Oppenheim,DCT}. For a smooth
finite-length sequence, its DCT is often energy-concentrated in
lower-indexed coefficients. Hence the DCT basis matrix
\begin{IEEEeqnarray}{rCl}
    [\mb Q_m]_{ij}
    ={q}_j\cos\left(\frac{\pi(2i-1)(j-1)}{2M}\right),
\end{IEEEeqnarray}
for $1\leq i\leq M$ and $1\leq j\leq m$, where $1\leq m\leq M$ and
\begin{IEEEeqnarray}{rCl}
q_j=\left\{
      \begin{array}{ll}
        \sqrt{{1}/{M}}, & j=1; \\
        \sqrt{{2}/{M}}, & j=2,\cdots,M.
      \end{array}
    \right.
\end{IEEEeqnarray}
is an excellent candidate RR basis for our channel estimation purpose.
Some comments on the predetermined basis selection are provided in
\cite{KLT,KLT2}.
\begin{remark}
As will be seen in the ensuing subsection, the proposed SSFC estimator
can be realized by performing an inverse DCT or KLT on the received
signal vector but the complexity of computing KLT and DCT are respectively
$\mathcal O(M^2)$ and $\mathcal O(M\log_2M)$. On the other hand, both
polynomial basis and DCT basis do not need the spatial correlation information
but DCT basis is computationally more efficient than the polynomial basis. \label{remark:KLT}
\end{remark}
\begin{remark}
The fact that the energy compaction efficiency of DCT is
near-optimal makes it the closest KLT approximation in the high
correlation regime among the following unitary transforms:
Walsh-Hadamard, Slant, Haar, and discrete Legendre transform, where
the last one is equivalent to a polynomial-based transform and is
slightly inferior to DCT in energy compaction capability.

The above claims have been verified in the context of image compression
\cite{KLT,KLT2}. In terms of RR MIMO channel representation, we show in {Section
\ref{section:simulation}} that, for the same modeling order $m$, the
DCT basis does outperform the polynomial basis in MU SSFC estimation
error regardless of the correlation level.
\label{remark:DCT_vs_poly}
\end{remark}

\subsection{SSFC Estimation}
We begin with the channel model (\ref{C_only}) and denote by
$\bm{\epsilon }^{(\textrm{I})}$ the modeling error. Let
${\gamma}=\sqrt{{\beta}}\|\mb{p}\|^2$ and assume for the moment
that LSFCs are known. Then
\begin{IEEEeqnarray}{rCl}
    \mb{Y}\mb{p}&=&
    {\sqrt{\beta}}
    \|\mb{p}\|^2\mb{h}+\mb{N}\mb{p}
     \notag\nonumber\\
    &=&\gamma\left(\mb{Q}_m^{(\textrm{I})}\mb{c}^{(\textrm{I})}+\bm{\epsilon }^{(\textrm{I})}\right)
     +\mb{N}\mb{p}
\end{IEEEeqnarray}
which brings about the following LS problem
\begin{IEEEeqnarray}{rCl}
\underset{{\mb{c}}}{\min}~~\left\|\mb{Y}\mb{p}-\gamma
\mb{Q}_{m}^{(\textrm{I})}\mb{c}^{(\textrm{I})}\right\|^2
\end{IEEEeqnarray}
The optimal solution can be shown as
\begin{IEEEeqnarray}{rCl}
\hat{\mb{c}}^{(\textrm{I})}
=\frac{1}{\gamma}\left(\hspace{-.1em}\mb{Q}_m^{(\textrm{I})}\hspace{-.1em}\right)^{\hspace{-.25em}H}\mb{Y}\mb{p}.
\label{eqn:36}
\end{IEEEeqnarray}
Replacing $\gamma$ by $\hat{\gamma}=\hat{\beta}^{\frac{1}{2}}\|\mb{p}\|^2$
for the case when LSFCs have to be estimated, we have
\begin{equation}
 \hat{\mb{h}}^{(\textrm{I})}
 = \mb{Q}_{m}^{(\textrm{I})}\hat{\mb{c}}^{(\textrm{I})}
 =\frac{1}{\gamma}\mb{Q}_{m}^{(\textrm{I})}
 \left(\hspace{-.1em}\mb{Q}_m^{(\textrm{I})}\hspace{-.1em}\right)^{\hspace{-.25em}H}\mb{Y}\mb{p}.
 \label{estimator:SSFC-2}
\end{equation}

On the other hand, if (\ref{RRApp}) is the channel model and
$\bm{\epsilon}^{(\textrm{II})}$ is the corresponding modeling error, then
\begin{IEEEeqnarray}{rCl}
    \mb{Y}\mb{p}&=&
    {\sqrt{\beta}}
    \|\mb{p}\|^2\mb{h}+\mb{N}\mb{p}
    \notag\nonumber\\
    &=&\gamma\left(\mb{W}(\phi)\mb{Q}_m^{(\textrm{II})}\mb{c}^{(\textrm{II})}+\bm{\epsilon}^{(\textrm{II})}\right)
     +\mb{N}\mb{p}
\end{IEEEeqnarray}
which suggests the LS formulation
\begin{IEEEeqnarray}{rCl}
\label{prob:28}
\min_{\phi,\mb{c}}&&~~\left\|\mb{Y}\mb{p}-\gamma
\mb{W}(\phi)\mb{Q}_{m}^{(\textrm{II})}\mb{c}^{(\textrm{II})}\right\|^2\nonumber\\
\mathrm{s.t.}&&~~\mb{W}(\phi)={\Diag}\left(\omega_1(\phi),
\cdots,\omega_M(\phi)\right), \nonumber\\
     &&~~~\omega_i(\phi)=\exp\left(-j2\pi\frac{(i-1)\xi}{\lambda}
     \sin\phi\right).
\end{IEEEeqnarray}
With $\mb{F}_m(\phi)\defeq\mb{W}(\phi)\mb{Q}_{m}^{(\textrm{II})}$ and
$\mb{A}^\dagger\defeq (\mb{A}^H\mb{A})^{-1}\mb{A}^H$, the optimal
solution to (\ref{prob:28}) is given as
\begin{IEEEeqnarray}{rCl}
\hat{\phi}
&=&\underset{\phi\in[-\frac{\pi}{2},\frac{\pi}{2}]}{\arg\max}~
\mb{p}^H
\mb{Y}^H\mb{F}_m(\phi)
\mb{F}_m^\dagger(\phi)\mb{Y}\mb{p}
\nonumber \\
&=&\underset{\phi\in[-\frac{\pi}{2},\frac{\pi}{2}]}{\arg\max}~
\left\|\left(\mb{W}(\phi)\mb{Q}_{m}^{(\textrm{II})}
\right)^H\mb{Y}\mb{p}\right\|^2.
\label{eqn:W_hati}\\
\hat{\mb{c}}^{(\textrm{II})}
&=&\frac{1}{\gamma}\mb{F}_m^{\dagger}(\hat{\phi})
\mb{Y}\mb{p}
=\frac{1}{\gamma}\left(\hspace{-.1em}\mb{Q}_m^{(\textrm{II})}\hspace{-.1em}\right)^{\hspace{-.25em}H}
\mb{W}^H(\hat{\phi})\mb{Y}\mb{p},
\label{c_hat}
\end{IEEEeqnarray}
When the true LSFCs are not available we use their estimates,
$\hat{\gamma}=\hat{\beta}^{\frac{1}{2}}\|\mb{p}\|^2$, and obtain
the SSFCs estimate
\begin{equation}
 \hat{\mb{h}}^{(\textrm{II})}=\mb{W}(\hat{\phi})\mb{Q}_{m}^{(\textrm{II})}
 \hat{\mb{c}}^{(\textrm{II})}.
 \label{h_hat}
\end{equation}
Both (\ref{eqn:W_hati}) and (\ref{c_hat}) require no matrix
inversion while $\hat{\phi}$ can be easily found by a simple line
search.

\subsection{Performance Analysis}
Assuming perfect LSFC knowledge, (\ref{estimator:SSFC-2}) becomes
\begin{IEEEeqnarray}{rCl}
\hat{\mb{h}}^{(\textrm{I})}
&=&\frac{1}{\gamma}\mb{Q}_{m}^{(\textrm{I})}\left(\hspace{-.1em}\mb{Q}_m^{(\textrm{I})}\hspace{-.1em}\right)^{\hspace{-.25em}H}
(\gamma\mb{h}+\mb{N}\mb{p})
\defeq\mathbb{E}\{\hat{\mb{h}}^{(\textrm{I})}\}+\bm{\nu}^{(\textrm{I})},~~
\end{IEEEeqnarray}
where $\bm{\nu}^{(\textrm{I})}=\frac{1}{\gamma}\mb{Q}_{m}^{(\textrm{I})}
\left(\hspace{-.1em}\mb{Q}_m^{(\textrm{I})}\hspace{-.1em}\right)^{\hspace{-.25em}H}
\mb{N}\mb{p}$,
and with (\ref{c_hat}) substituting into it,
(\ref{h_hat}) becomes
\begin{IEEEeqnarray}{rCl}
\hat{\mb{h}}^{(\textrm{II})}&=&\frac{1}{\gamma}\mb{W}(\hat{\phi})
\mb{Q}_{m}^{(\textrm{II})}\left(\hspace{-.1em}\mb{Q}_m^{(\textrm{II})}\hspace{-.1em}\right)^{\hspace{-.25em}H}
\mb{W}^H(\hat{\phi})\mb{Y}\mb{p}\notag\\
&=&\frac{1}{\gamma}\mb{W}(\hat{\phi})
\mb{Q}_{m}^{(\textrm{II})}\left(\hspace{-.1em}\mb{Q}_m^{(\textrm{II})}\hspace{-.1em}\right)^{\hspace{-.25em}H}
\mb{W}^H(\hat{\phi})(\gamma\mb{h}+\mb{N}\mb{p})~~~~\nonumber\\
&\defeq&\mathbb{E}\{\hat{\mb{h}}^{(\textrm{II})}\}+\bm{\nu}^{(\textrm{II})},
\label{eqn:h_biased}
\end{IEEEeqnarray}
where 
$\bm{\nu}^{(\textrm{II})}=\frac{1}{\gamma}\mb{W}(\hat{\phi})
\mb{Q}_{m}^{(\textrm{II})}\left(\hspace{-.1em}\mb{Q}_m^{(\textrm{II})}\hspace{-.1em}\right)^{\hspace{-.25em}H}
\mb{W}^H(\hat{\phi})\mb{N}\mb{p}$.
We denote by
$\MSE_m(\cdot)$ the MSE of the enclosed SSFC estimate with modeling order $m$ and decompose it into
\begin{IEEEeqnarray}{rCl}
\MSE_m(\hat{\mb{h}})&=&
\mathbb{E}\left\{\left\|\hat{\mb{h}}-\mb{h} \right\|^2\right\}\notag\\
&=&\underbrace{\mathbb{E}\left\{
\left\|\hat{\mb{h}}-\mathbb{E}\{\hat{\mb{h}}\}
\right\|^2\right\}}_{\defeq\var\{{\hat{\mb{h}}}\}}+
\underbrace{\mathbb{E}\left\{
\left\|\mathbb{E}\{\hat{\mb{h}}\}-\mb{h}
\right\|^2\right\}}_{\defeq b({\hat{\mb{h}}})}\vspace{-1.2em}\notag\\
\label{eqn:23}
\end{IEEEeqnarray}
$\var\{{\hat{\mb{h}}}\}$ and $b({\hat{\mb{h}}})$ represent respectively
the variance and bias of estimator $\hat{\mb{h}}$.
For these two error terms we prove in \ref{app:pf_MSE} that
\begin{theorem}
\label{thm:MSE}
For SSFC estimators $\hat{\mb{h}}^{(\textrm{I})}$ and $\hat{\mb{h}}^{(\textrm{II})}$,
\begin{IEEEeqnarray}{rCl}
\var\{\hat{\mb{h}}^{(\textrm{I})}\}=
\var\{\hat{\mb{h}}^{(\textrm{II})}\}
&=&\frac{m}{\beta\|\mb{p}\|^2},\notag
\label{eqn:var_term}
\end{IEEEeqnarray}
and
\begin{IEEEeqnarray}{rCl}
b(\hat{\mb{h}}^{(\textrm{I})})
&=&\tr\left(\mb{D}_m\left(\hspace{-.1em}\mb{Q}^{(\textrm{I})}\hspace{-.1em}\right)^{\hspace{-.25em}H}
\mb{\Phi}\mb{Q}^{(\textrm{I})}\right)\notag\\
b(\hat{\mb{h}}^{(\textrm{II})})
&=&\tr\left(\mb{D}_m\left(\hspace{-.1em}\mb{Q}^{(\textrm{II})}\hspace{-.1em}\right)^{\hspace{-.25em}H}
\mb{W}^H(\hat{\phi})\mb{\Phi}\mb{W}(\hat{\phi})\mb{Q}^{(\textrm{II})}\right)\notag
\label{eqn:bias_term}
\end{IEEEeqnarray}
where $\mb{D}_m=\Diag\left(\left[\mb{0}_{1\times
m}~\mb{1}_{1\times(M-m)}\right]^T\right)$.
\end{theorem}

\begin{remark} 
If full-rank model, $m=M$, is used, then $\mb{Q}_{m}
\mb{Q}_{m}^H=\mb{I}_M$. It is easy to check ${\bf D}_m={\bf 0}_M$
and to see from (\ref{estimator:SSFC-2}) and (\ref{h_hat}) that the proposed SSFC estimators
$\hat{\mb{h}}^{(\textrm{I})}$ and $\hat{\mb{h}}_k^{(\textrm{II})}$
equivalent to the conventional unbiased LS estimator
\cite{Opt_pilot}
\begin{IEEEeqnarray}{rCl}
\hat{\mb{h}}=\frac{1}{\gamma}\mb{Y}\mb{p}.
\label{eqn:convetional_LS}
\end{IEEEeqnarray}
\label{prop:conventional_LS}
\end{remark}

\section{Simulation Results}
\label{section:simulation}

In this section, we investigate the performance of the proposed
estimators via simulation with a standardized channel model--SCM--whose spatial correlation at the BS is related to
AoA distribution and antenna spacings \cite{SCM}. In addition, the
environment surrounding a user is of rich scattering with AoDs
uniformly distributed in $[-\pi, \pi)$ making spatial correlation between MSs negligible. This setting accurately describes the environment
where the BS with large-scale antenna array are mounted on an
elevated tower or building. We assume that there are $8$ users located randomly in a circular cell of radius $R$ with their mean AoAs uniformly distributed within $[-60^\circ,60^\circ]$. The other
simulation parameters are listed in Table \ref{tab:sim}. We define
average received signal-to-noise power ratio as
$\mathrm{SNR}\defeq\beta\|\mb{p}\|^2/T$ and
normalized MSE (NMSE) as the
MSE between the true and estimated vectors normalized by the
former's dimension and entry variance. Note that for LSFC estimation,
\[
\mathrm{NMSE}(\hat{\beta}_{\textrm{dB}})\defeq
\mathbb{E}\left\{\left(10\log{\hat{\beta}}/{\beta}\right)^2\right\}/
\var\{10\log\beta\}
\]
instead of $\mathrm{NMSE}(\hat{\beta})$
is considered in this section.
\begin{table}[t]
 \caption{Simulation parameters}
 \centering
 \tabcolsep 0.15in
 \label{tab:sim}
 \begin{tabular}{|l|l|}
 \hline
 \textit{Parameter} & \textit{Value}\\ \hline
 Cell radius $R$ & $100$ meters \\ \hline
 Pathloss exponent $\alpha$ & $3$ \\ \hline
 Shadowing standard deviation $\sigma_s$ & $10$ dB \\ \hline
 Number of BS antennas $M$ & $100$ \\ \hline
 BS antenna spacing $\xi$ & 0.5$\lambda$ \\ \hline
 Number of MSs $K$ & $8$ \\ \hline
 Number of path in SCM & $1$ \\ \hline
 Number of subpath in SCM & $20$ \\ \hline
 \end{tabular}
\end{table}

\begin{figure}
    \centering
        \includegraphics[width=3.6in]{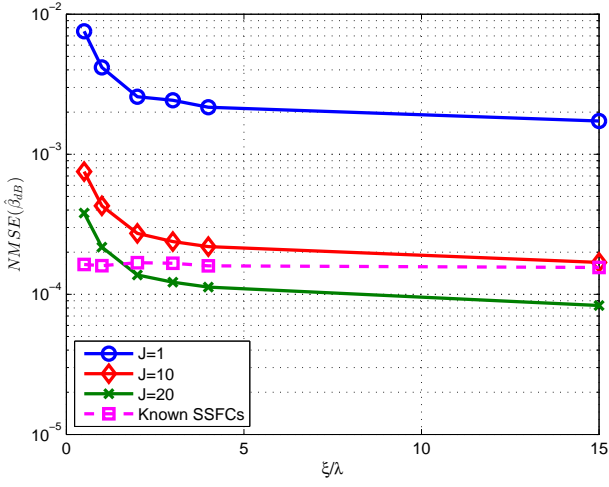}
        \caption{Effect of antenna spacing on the
        conventional LSFC estimator (with perfect SSFC
        knowledge) and proposed LSFC estimators using one or multiple blocks; $\mathrm{AS}=15^\circ$, $\SNR=10$ dB.}
 \label{fig:compare_beta}
\end{figure}
In Fig. \ref{fig:compare_beta} we compare the performance of the
proposed LSFC estimators (\ref{ALS}) and (\ref{ALS_J}) with that of a conventional LS
estimator \cite[Ch. 8]{Kay}
\begin{IEEEeqnarray}{rCl}
\widehat{\sqrt{\bm\beta}}=(\mb{A}^H\mb{A})^{-1}\mb{A}^H
\mathrm{vec}(\mb{Y}),
\end{IEEEeqnarray}
where $\mb{A}=\left(\mb{1}_T\otimes \mb{H}\right) \odot\left(
\mb{P}^T\otimes \mb{1}_M\right)$. As opposed to our proposal, the
conventional estimator needs to know SSFCs beforehand, hence
full knowledge of SSFCs is assumed for the latter. As can be seen, as antenna spacing increases, the channel
decorrelates and thus the estimation error due to spatial
correlation decreases; this verifies \textit{Theorem
\ref{prop:Mean_phi}}. Figure \ref{fig:compare_beta} also shows that our proposed estimator attains the performance of the conventional one (using one block) when channel correlation decreases to $0$ with $J=10$ training blocks and outperforms the conventional when $J=20$. This suggests that we can have good LSFC estimates even when the SSFCs are not available due to the advantage of the noise reduction effect that massive MIMO systems have offered.
\begin{figure}
    \centering
        \includegraphics[width=3.6in]{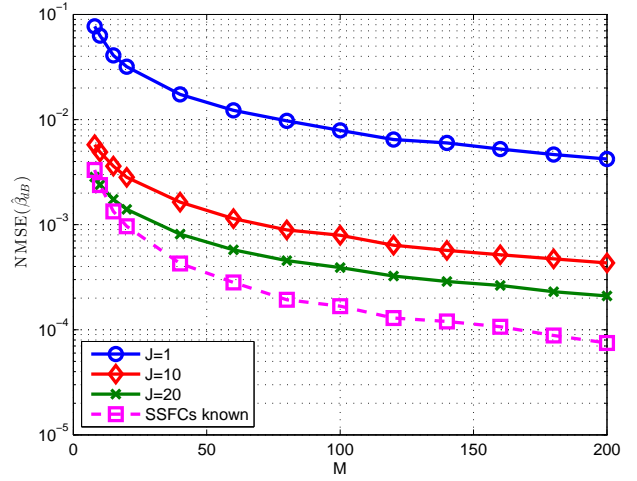}
        \caption{Large-system performance of the proposed LSFC estimators using one or more pilot blocks and that of the conventional estimator with perfect SSFC knowledge; $\mathrm{AS}=15^\circ$, $\SNR=10$ dB.}
 \label{fig:error_number}
\end{figure}
On the other hand, Fig. \ref{fig:error_number} illustrates the effect
of massive antennas to the MSE. Owing to the fact that we have assumed
perfect SSFC knowledge for the conventional LSFC estimator, MSE
decreases with increasing sample amount as $M$ increases.
Unlike the conventional, the amount of known information does not
grow with $M$ for the proposed LSFC estimator. However, channel hardening effect
becomes more serious and thus improves the estimator accuracy.

\begin{figure}
    \centering
        \includegraphics[width=3.6in]{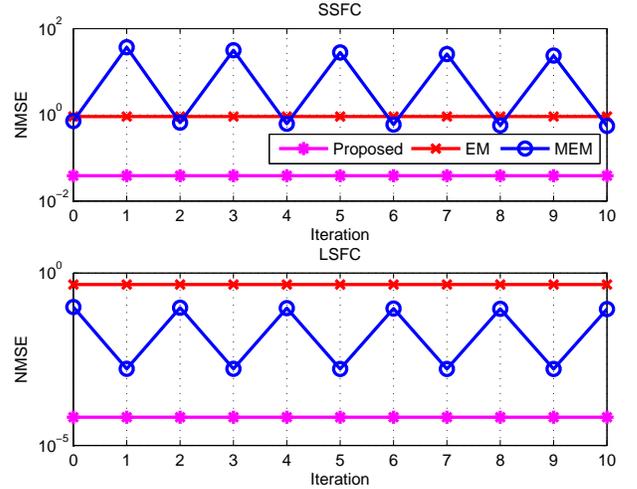}
        \caption{MSE performance of two EM-based joint LSFC/SSFC estimators versus their iteration numbers (which means initialization for value $0$); $\mathrm{AS}=7.2^\circ$, $\SNR=10$ dB. MSE of the proposed full-order SSFC and LSFC estimators is also plotted.}

 \label{fig:EM_10dB}
\end{figure}
The performance of an EM-based joint LSFC and SSFC estimation is
shown in Fig. \ref{fig:EM_10dB}.
When the LSFC and SSFCs of an MS-BS link are all unknown and
channel hardening effect in massive MIMO is disregarded,
the coupling nature of the LSFC and SSFCs suggests
EM algorithm be applied to derive their estimator.
The EM-based joint LSFC and SSFC estimation is detailed as follows with $\sqrt{\boldsymbol{\beta}}\defeq[\sqrt{\beta_1},\cdots,\sqrt{\beta_K}]^T$:
\begin{enumerate}
\item(Initialization) Initialize $\widehat{\sqrt{\boldsymbol{\beta}}}
=\mathbb{E}\{\sqrt{\bm{\beta}}\}$.
\item(Updating SSFC Estimates) Let $\hat{\bm{\beta}}=
\left(\widehat{\sqrt{\boldsymbol{\beta}}}\right)^2$. Calculate
\begin{IEEEeqnarray}{rCl}
\mathrm{vec}(\hat{\mathbf{H}})=\Diag\Big((\mb{\Phi}_1+\|
\mathbf{p}_1\|^2\hat{\beta}_1\mathbf{I}_M)^{-1}, \cdots, ~~&&\nonumber\\
(\mb{\Phi}_K+\|\mathbf{p}_K\|^2\hat{\beta}_K\mathbf{I}_M)^{-1} \Big)&&\notag\\
\cdot\left(\mb{D}_{\hat{\bm{\beta}}}^{\frac{1}{2}}\mb{P}^\star\otimes \mathbf{I}_M\right)&&\mathrm{vec}(\mathbf{Y}).~~~\notag
\end{IEEEeqnarray}
\item(Updating LSFC Estimates) Calculate
\begin{IEEEeqnarray}{rCl}
\widehat{\sqrt{\boldsymbol{\beta}}}
=\mathbb{E}\left\{\sqrt{\boldsymbol{\beta}}\right\}
&+&\left(\mathbf{C}^{-1}
+\mathbf{A}^H\mathbf{A}\right)^{-1}\nonumber\\
&&~\cdot\mathbf{A}^H\left(\mathrm{vec}(\mathbf{Y})-\mathbf{A}\:\mathbb{E}\left\{\sqrt{\boldsymbol{\beta}}\right\}\right)
\notag
\end{IEEEeqnarray}
where $\mb{A}=\left(\mb{1}_T\otimes \hat{\mb{H}}\right) \odot\left(
\mb{P}^T\otimes \mb{1}_M\right)$ and $\mb{C}$ is the covariance matrix of $\bm{\beta}$.
\item(Recursion) Go to Step 2); or terminate and output $\hat{\mb{H}}$ and
$\hat{\bm{\beta}}=
\left(\widehat{\sqrt{\boldsymbol{\beta}}}\right)^2$ if convergence is achieved.
\end{enumerate}
Moreover, since
\begin{IEEEeqnarray}{rCl}
\mathbf{A}^H\mathbf{A}&=&(\mathbf{H}^H\mathbf{H})\odot(\mathbf{P}^\star\mathbf{P}^T)
\notag\\
&\as& \Diag\left(M\|\mathbf{p}_1\|^2,\cdots, M\|\mathbf{p}_K\|^2\right)\notag
\end{IEEEeqnarray}
a modified EM (MEM) algorithm is obtained by replacing $\mathbf{A}^H\mathbf{A}$
with $\Diag\left(M\|\mathbf{p}_1\|^2,\cdots, M\|\mathbf{p}_K\|^2\right)$ in Step 3).
Clearly, the proposed LSFC/SSFC-decoupled estimator outperforms the EM-based ones significantly. While the former requires no recursion and thus saves computation, the modified EM algorithm cannot converge in a limited number of iterations.

\begin{figure}
    \centering
        \includegraphics[width=3.6in]{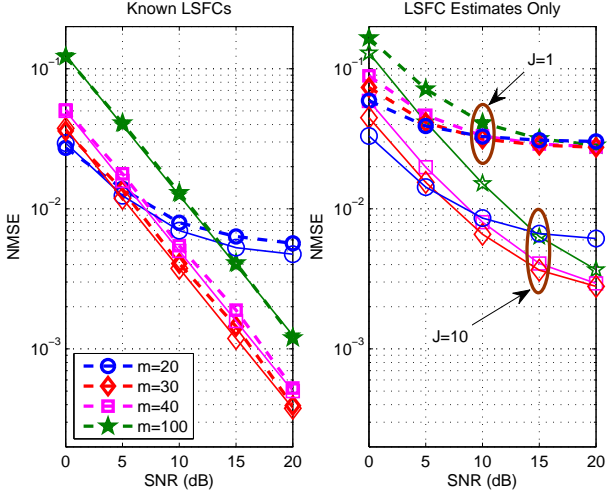}
        \caption{MSE of the proposed RR SSFC estimator using polynomial bases of various ranks;
        $\mathrm{AS}=7.2^\circ$. Theoretical MSE (\ref{eqn:23}) with known mean AoA is given (in solid lines) in the left plot. The right plot shows the SSFC estimation performance with LSFC estimated from $J=1$ or $10$ blocks.}
 \label{fig:model_order_72_poly}
\end{figure}

\begin{figure}
    \centering
        \includegraphics[width=3.6in]{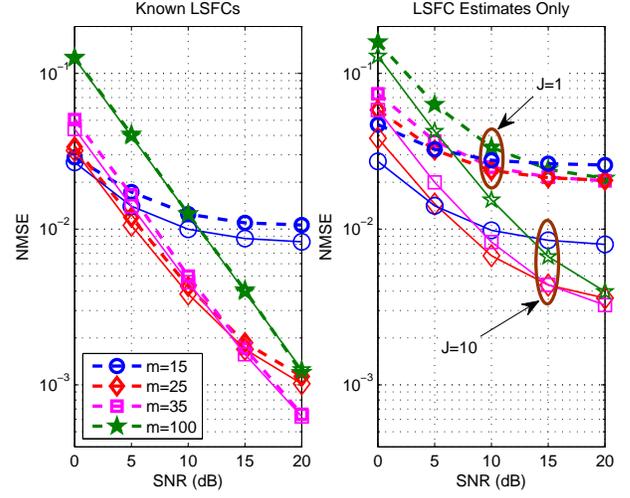}
        \caption{MSE of the proposed DCT-based SSFC estimator of various ranks; $\mathrm{AS}=7.2^\circ$. The right plot shows the SSFC estimation performance with LSFC estimated from $J=1$ or $10$ blocks, whereas theoretical MSE (\ref{eqn:23}) having known AoAs  is given in the left plot (in solid lines). }
 \label{fig:model_order_72_DCT}
\end{figure}

\begin{figure}
    \centering
        \includegraphics[width=3.6in]{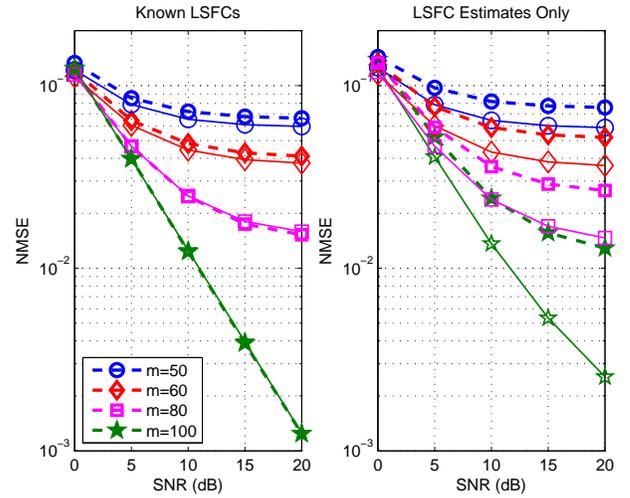}
        \caption{MSE of the proposed polynomial-based SSFC estimator with various ranks;
        $\mathrm{AS}=15^\circ$. Theoretical MSE (\ref{eqn:23}) with known mean AoA is given (in solid lines) in the left plot, while the MSE with LSFC estimated from $J=1$ and $10$ blocks in dashed and solid lines, respectively, in the right.}
 \label{fig:model_order_15_poly}
\end{figure}

\begin{figure}
    \centering
        \includegraphics[width=3.6in]{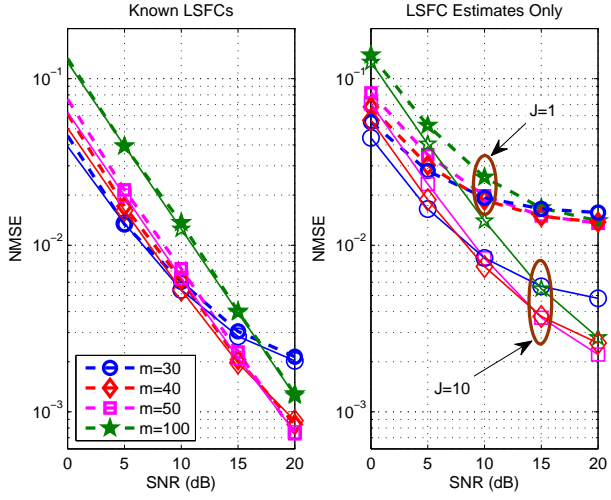}
        \caption{MSE of the proposed RR SSFC estimator using DCT bases of various ranks;
        $\mathrm{AS}=15^\circ$. Theoretical MSE (\ref{eqn:23}) with perfect mean AoA knowledge is given (in solid lines) in the left plot. The right plot shows the SSFC estimation performance with LSFC estimated from $J=1$ or $10$ blocks.}
 \label{fig:model_order_15_DCT}
\end{figure}

In addition, we study the SSFC estimation performance of $\hat{\mb{h}}^{(\textrm{II})}$
with respect to modeling order and basis matrix with estimated or
perfectly-known $\bm{\beta}$ in Figs.  \ref{fig:model_order_72_poly}--\ref{fig:model_order_15_DCT}.
Since the spatial
correlation increases with reducing AS, the spatial waveform of an MS over the array is
anticipated to be smoother. As a result, when AS is comparatively small, due to over-modeling the channel vector the estimation
performance not only cannot be improved, but also
may be degraded. This is because the amount of available information does not grow with
that of the parameters required to be obtained. As can be observed in Fig. \ref{fig:model_order_72_poly},
when LSFCs are perfectly known, the estimation accuracy with polynomial basis degrades
as modeling order increases from $20$ to $100$ for $\mathrm{SNR}<5$ dB. Besides,
the optimal modeling order increases with $\mathrm{SNR}$, e.g., optimal order at
$\mathrm{SNR}=0$ and $5$ dB are respectively $20$ and $30$. Such result is observed
due to the fact that MSE is noise-limited in the low SNR regime, while the importance
of modeling error becomes more pronounced for high SNRs. Similar trend is also observed
with DCT basis in Fig. \ref{fig:model_order_72_DCT}.

On the other hand, when $\mathrm{AS}$ increases to $15^\circ$, the spatial correlation
decreases and spatial waveforms roughen. As can be seen in Fig. \ref{fig:model_order_15_poly},
the SSFC estimator fails to capture these waveforms by using only some low-degree polynomials,
and thus in this scenario, full modeling order ($M=100$) is required to achieve best performance
for any SNR. However, DCT basis is still appropriate for parameter reduction due to the near-optimal energy-compacting
capability od DCT as described \textit{Remark \ref{remark:DCT_vs_poly}}. In Fig. \ref{fig:model_order_15_DCT}, we observe that when LSFCs are perfectly estimated, a modeling
order of $30$ has the minimal MSE for $\mathrm{SNR}<9$ dB. Moreover, we can see from the right plots of Figs. \ref{fig:model_order_72_poly}--\ref{fig:model_order_15_DCT} that the estimation of SSFCs requires accurate LSFC estimates. Such estimates can be easily obtained with multiple pilot blocks due to LSFCs' slowly-varying characteristics.

\begin{figure}
    \centering
        \includegraphics[width=3.6in]{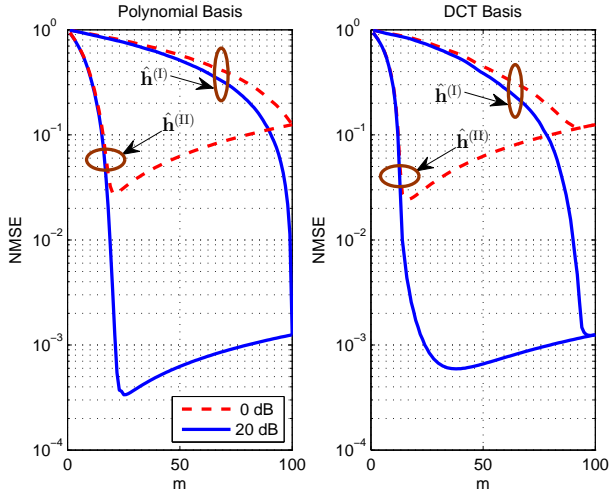}
        \caption{Theoretical MSE of
        $\hat{\mb{h}}^{(\textrm{I})}$ and $\hat{\mb{h}}^{(\textrm{II})}$ versus
        modeling order and $\SNR$;
        $\mathrm{AS}=7.2^\circ$.}
 \label{fig:WC_vs_C_compare_AS72}
\end{figure}

\begin{figure}
    \centering
        \includegraphics[width=3.6in]{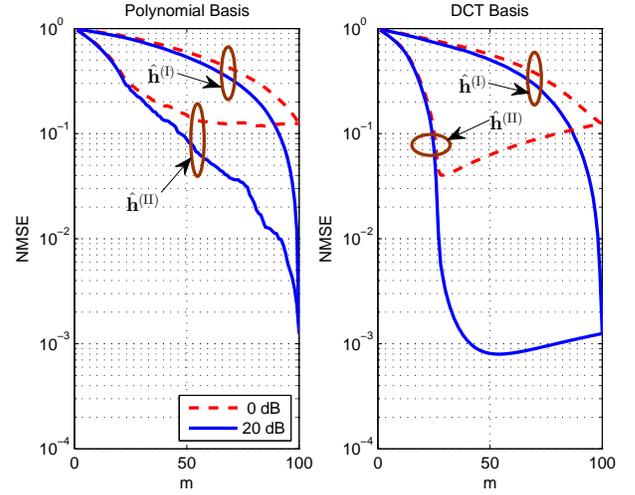}
        \caption{Theoretical MSE of $\hat{\mb{h}}^{(\textrm{I})}$ and $\hat{\mb{h}}^{(\textrm{II})}$
        versus modeling order and $\SNR$; $\mathrm{AS}=15^\circ$.}
 \label{fig:WC_vs_C_compare_AS15}
\end{figure}

Figs. \ref{fig:WC_vs_C_compare_AS72} and \ref{fig:WC_vs_C_compare_AS15} plot the theoretical
MSE (\ref{eqn:23}) of the proposed two SSFC estimators with respect to the modeling order $m$
and compare the performance between $\hat{\mb{h}}^{(\textrm{I})}$ and $\hat{\mb{h}}^{(\textrm{II})}$
using a same basis matrix.
While these plots are able to suggest some basis selecting guidance in different scenarios,
they also substantiate the results shown in Figs. \ref{fig:model_order_72_poly}--\ref{fig:model_order_15_DCT}.
First, we investigate the results of $\hat{\mb{h}}^{(\textrm{II})}$. When $\mathrm{AS}=7.2^\circ$, $\SNR=20$ dB,
and, the polynomial basis is capable of rendering a better estimation performance with greater parameter number
reduction than the DCT basis can provide, i.e., the optimal modeling order for the polynomial and DCT basis are
respectively $25$ and $38$. However, when $\SNR$ reduces to $0$ dB, the DCT basis achieves the lowest MSE with
$m=16$ as compared to $m=21$ of the polynomial basis. In spite of the effectiveness of the polynomial basis
when $\mathrm{AS}=7.2^\circ$, it fails to represent channel with only a few low degree polynomials when
$\mathrm{AS}=15^\circ$ and requires almost full order. On the other hand, the DCT basis is still applicable
with optimal order $m=29$ and $54$ for $\SNR=0$ and $20$ dB, respectively.

As for $\hat{\mb{h}}^{(\textrm{I})}$, the fact that in the given scenarios its MSE performance is significantly
worse than that of $\hat{\mb{h}}^{(\textrm{II})}$ for any $m$ and that the optimal order are all close to $M=100$
justifies our preference to $\hat{\mb{h}}^{(\textrm{II})}$. Recall that both $\hat{\mb{h}}^{(\textrm{I})}$ and
$\hat{\mb{h}}^{(\textrm{II})}$ degenerate to the conventional LS estimator (\ref{eqn:convetional_LS}) if $m=100$,
thus they all have the same performance regardless of the basis chosen. Although, applying the polynomial basis
when $\mathrm{AS}=15^\circ$ causes performance inferior to that of the conventional LS estimator,
$\hat{\mb{h}}^{(\textrm{I})}$ offers direct performance-complexity trade-off.


\section{Conclusion}
\label{section:conclusion} Taking advantage of the noise reduction effect and the
large number of samples available in a massive MIMO system, we propose a novel LSFC
estimator for both spatially-correlated and uncorrelated channels. This estimator
is easily extendable to the case when multiple pilot blocks become available.
The estimator is of low complexity, requires no prior knowledge of SSFCs and spatial
correlation, and yields asymptotically diminishing MSE when the number of BS antennas
becomes large enough.

Using the estimated LSFCs, we present an algorithm which performs joint SSFC and mean
AoA estimation with rank-reduced channel model. The simultaneous mean AoA estimation
not only offer useful information for downlink beamforming but also improves the SSFC
estimator's performance. A closed-form MSE expression for the proposed SSFC estimator
is derived to investigate the rank reduction efficiency.

Moreover, we compare some candidate bases for RR representation and examine their effects
on SSFC estimation. We show that the DCT basis is an excellent choice due to its low
computing complexity and, more importantly to its energy compaction capability.
Numerical results confirm the effectiveness of the proposed estimator and verify that
the optimal modeling order is a function of AS and SNR.

\appendices
\renewcommand{\thesection}{Appendix \Alph{section}}
\renewcommand{\theequation}{\Alph{section}.\arabic{equation}}
\renewcommand{\thelemma}{\Alph{section}.\arabic{lemma}}
\setcounter{equation}{0}
\setcounter{lemma}{0}
\section{Proof of {Theorem \ref{prop:Mean_phi}}}
\label{pf:Thm_1} \setcounter{equation}{0}

\textit{Lemma \ref{lemma:trace}} implies that if
\begin{IEEEeqnarray}{rCl}
\underset{{M\rightarrow\infty}}{\limsup}\underset{{1\leq i,j\leq K}}
{\sup}\|\mb{\Phi}_{i}^{\frac{1}{2}}\mb{\Phi}_{j}^{\frac{1}{2}}\|_2<\infty,
\label{cond:1}
\end{IEEEeqnarray}
we have
\begin{IEEEeqnarray}{rCl}
&&\hspace{-1em}\frac{1}{M}\tilde{\mb{H}}^H\mb{A}\tilde{\mb{H}} \nonumber \\
&&=\frac{1}{M}\left[\begin{array}{ccc}
            \tilde{\mb{h}}_{1}^H\mb{\Phi}_1^{\frac{1}{2}}\mb{\Phi}_1^{\frac{1}{2}}\tilde{\mb{h}}_{1} & \cdots
            &\tilde{\mb{h}}_{1}^H\mb{\Phi}_1^{\frac{1}{2}}\mb{\Phi}_K^{\frac{1}{2}}\tilde{\mb{h}}_{K}\\
            \vdots & \ddots & \vdots \\
            \tilde{\mb{h}}_{K}^H\mb{\Phi}_K^{\frac{1}{2}}\mb{\Phi}_1^{\frac{1}{2}}
            \tilde{\mb{h}}_{1} & \cdots &\tilde{\mb{h}}_{K}^H\mb{\Phi}_K^{\frac{1}{2}}\mb{\Phi}_K^{\frac{1}{2}}\tilde{\mb{h}}_{K}
            \end{array}\right] \as\mb{I}_K, \notag\\
\end{IEEEeqnarray}
and if 
$\underset{{M\rightarrow\infty}}{\limsup}\underset{{1\leq i\leq
K}}{\sup}\|\mb{\Phi}_{i}^{\frac{1}{2}}\|_2<\infty$,
\begin{IEEEeqnarray}{rCl}
\underset{{M\rightarrow\infty}}{\limsup}\underset{{1\leq
i\leq K}}{\sup}\|\mb{\Phi}_{i}\|_2&<&\infty, \label{cond:2}
\end{IEEEeqnarray}
then
\begin{IEEEeqnarray}{rCl}
&&\hspace{-1em}\frac{1}{M}\tilde{\mb{H}}^H\tilde{\mb{\Phi}}^H\mb{N} \nonumber\\
&&=\frac{1}{M}\left[\begin{array}{ccc}
            \tilde{\mb{h}}_{1}^H\mb{\Phi}_1^{\frac{1}{2}}\mb{n}_1 & \cdots &\tilde{\mb{h}}_{1}^H\mb{\Phi}_1^{\frac{1}{2}}\mb{n}_K\\
            \vdots & \ddots & \vdots \\
            \tilde{\mb{h}}_{K}^H\mb{\Phi}_K^{\frac{1}{2}}\mb{n}_1 & \cdots &\tilde{\mb{h}}_{K}^H\mb{\Phi}_K^{\frac{1}{2}}\mb{n}_K
            \end{array}\right] \as\mb{0}_{K\times T}. \notag
\end{IEEEeqnarray}
Note that \textit{Assumption \ref{assump:1}} is equivalent to
(\ref{cond:2}) and implies (\ref{cond:1}) since if $\forall~ i$, $\|\mb{\Phi}_{i}^{\frac{1}{2}}\|_2<\infty$, then
\begin{IEEEeqnarray}{rCl}
\|\mb{\Phi}_{i}^{\frac{1}{2}}\mb{\Phi}_{j}^{\frac{1}{2}}\|_2
{\leq}\|\mb{\Phi}_{i}^{\frac{1}{2}}\|_2\|\mb{\Phi}_{j}^{\frac{1}{2}}
\|_2&<&\infty,~\forall ~1\leq i,j \leq K. 
 \notag
\end{IEEEeqnarray}

\section{Proof of Theorem \ref{thm:MSE}}
\label{app:pf_MSE}
\setcounter{equation}{0}
\setcounter{lemma}{0}
In the following, we derive the variance and bias terms of the MSE of
$\hat{\mb{h}}^{(\textrm{II})}$. Those of the MSE of $\hat{\mb{h}}^{(\textrm{I})}$
can be similarly obtained. We start with the derivation of the variance term.
\begin{IEEEeqnarray}{rCl}
\var\{\hat{\mb{h}}^{(\textrm{II})}\}
&=&\mathbb{E}\left\{\left\|\hat{\mb{h}}^{(\textrm{II})}-\mathbb{E}
\{\hat{\mb{h}}^{(\textrm{II})}\}\right\|^2\right\}~~~~\nonumber\\
&=&\frac{1}{\gamma^2}\mb{p}^H\mathbb{E}\hspace{-.25em}\left\{
\hspace{-.25em}\mb{N}^H\mb{W}(\hat{\phi})
\mb{Q}_{m}^{(\textrm{II})}\left(\hspace{-.1em}\mb{Q}_m^{(\textrm{II})}\hspace{-.1em}\right)^{\hspace{-.25em}H}\hspace{-.3em}
\mb{W}^H(\hat{\phi})\mb{N}\hspace{-.25em}\right\}\hspace{-.25em}\mb{p}\nonumber\\
&=&\frac{1}{\gamma^2}\mb{p}^H\mathrm{tr}\hspace{-.25em}\left(\hspace{-.25em}\mb{W}(\hat{\phi})
\mb{Q}_{m}^{(\textrm{II})}\left(\hspace{-.1em}\mb{Q}_m^{(\textrm{II})}\hspace{-.1em}\right)^{\hspace{-.25em}H}\hspace{-.3em}
\mb{W}^H(\hat{\phi})\hspace{-.25em}\right)\hspace{-.25em}\mb{p}\label{eq:C1}\\
&{=}&\frac{1}{\gamma^2}\mb{p}^H(m\mb{I}_K)\mb{p}
=\frac{m}{\beta\|\mb{p}\|^2}\notag
\end{IEEEeqnarray}
where we have invoked the relation
\begin{IEEEeqnarray}{rCl}
\mathbb{E}\left\{\mb{N}^H\mb{X}\mb{N}\right\}
&=&\sum_{i=1}^{M}\sum_{j=1}^{M}x_{ij}\mathbb{E}\left\{\mb{n}_i\mb{n}_j^H
\right\} \nonumber\\
&=&\sum_{i=1}^{M}x_{ii}\mathbb{E}\left\{\mb{n}_i\mb{n}_i^H\right\}
=\mathrm{tr}(\mb{X})\mb{I}_K~~~
\end{IEEEeqnarray}
with white noise $\mb{N}=\left[\mb{n}_1,\cdots,\mb{n}_M\right]^H$
for any square matrix $\mb{X}=[x_{ij}]$. For the bias term, we have
\begin{IEEEeqnarray}{rCl}
b(\hat{\mb{h}}^{(\textrm{II})})&=&\mathbb{E}\left\{\left
\|\mathbb{E}\{\hat{\mb{h}}^{(\textrm{II})}\}-\mb{h}
\right\|^2\right\} \nonumber\\
&=&\mathbb{E}\left\{\mb{h}^H\hspace{-.25em}\left(\hspace{-.25em}\mb{W}(\hat{\phi})
\mb{Q}_{m}^{(\textrm{II})}\left(\hspace{-.1em}\mb{Q}_m^{(\textrm{II})}\hspace{-.1em}\right)^{\hspace{-.25em}H}
\mb{W}^H(\hat{\phi})-\mb{I}_M\hspace{-.25em}\right)^{\hspace{-.2em}2}
\hspace{-.15em}\mb{h}\right\}\nonumber\\
&=&\tr\left(\hspace{-.25em}\left(\mb{W}(\hat{\phi})
\mb{Q}_{m}^{(\textrm{II})}\left(\hspace{-.1em}\mb{Q}_m^{(\textrm{II})}\hspace{-.1em}\right)^{\hspace{-.25em}H}
\mb{W}^H(\hat{\phi})-\mb{I}_M\hspace{-.25em}\right)^{\hspace{-.2em}2}
\hspace{-.15em}\mb{\Phi}\hspace{-.25em}\right)\vspace{-1em}\nonumber\\
&&\hspace{1.7em}\underbrace{\hspace{15.5em}}_{\defeq\mb{A}_1}\notag
\end{IEEEeqnarray}
Since
\begin{IEEEeqnarray}{rCl}
\mb{Q}_{m}^{(\textrm{II})}\hspace{-.25em}
\left(\mb{Q}_{m}^{(\textrm{II})}\right)^{\hspace{-.25em}H}\hspace{-.5em}
&=&\hspace{-.4em}\mb{Q}^{(\textrm{II})}\hspace{-.25em}
\left[\hspace{-.3em}\begin{array}{c}
            \mb{I}_{m}\\
            \mb{0}_{(M-m)\times m}
            \end{array}\hspace{-.3em}\right]\hspace{-.25em}
\left[ 
            \mb{I}_{m} ~ \mb{0}_{(M-m)\times m}
       \right]\hspace{-.25em}
   \left(\hspace{-.1em}\mb{Q}^{(\textrm{II})}\hspace{-.1em}\right)^{\hspace{-.25em}H}
\nonumber\\
&\defeq&\hspace{-.4em}\mb{Q}^{(\textrm{II})}(\mb{I}_M-\mb{D}_m)
\left(\hspace{-.1em}\mb{Q}^{(\textrm{II})}\hspace{-.1em}\right)^{\hspace{-.25em}H}
\end{IEEEeqnarray}
where $\mb{D}_m=\Diag\left(\left[\mb{0}_{1\times m}~\mb{1}_{1\times(M-m)}\right]^T\right)$,
we have $\mb{A}_1=\mb{W}(\hat{\phi})\mb{Q}^{(\textrm{II})}
\mb{D}_m\left(\hspace{-.1em}\mb{Q}^{(\textrm{II})}\hspace{-.1em}\right)^{\hspace{-.25em}H}\mb{W}^H(\hat{\phi})$
and
\begin{IEEEeqnarray}{rCl}
b(\hat{\mb{h}}^{(\textrm{II})})
&=&\tr\left(\mb{W}(\hat{\phi})\mb{Q}^{(\textrm{II})}\mb{D}_m
\left(\hspace{-.1em}\mb{Q}^{(\textrm{II})}\hspace{-.1em}\right)^{\hspace{-.25em}H}
\mb{W}^H(\hat{\phi})\mb{\Phi}\right)\label{eqn:46}\nonumber\\
&=&\tr\left(\mb{D}_m\left(\hspace{-.1em}\mb{Q}^{(\textrm{II})}\hspace{-.1em}\right)^{\hspace{-.25em}H}
\mb{W}^H(\hat{\phi})\mb{\Phi}\mb{W}(\hat{\phi})\mb{Q}^{(\textrm{II})}\right).~~~~
\end{IEEEeqnarray}

\section{On the Spatial Correlation}
\label{app:sp_corr}
\setcounter{equation}{0}
\setcounter{lemma}{0}
Following \cite{SCM_v}, we express the spatial correlation between two antenna elements
in an array with arbitrary configuration as
\begin{IEEEeqnarray}{rCl}
  [\mb{\Phi}]_{ij}&=&\mathbb{E}\{h_i h_j^*\}\notag\\
   &=&\int_{-\pi}^{\pi}p(\theta)\exp\left(j\mb{k}^T(\theta)
   (\mb{u}_i-\mb{u}_j)\right)\mathrm{d}\theta
\end{IEEEeqnarray}
where $p(\theta)$ is the probability density function of AoA, $\mb{k}(\theta)=
-\frac{2\pi}{\lambda}[\cos(\theta)~\sin(\theta)]^T$, and $\mb{u}_i=[u_{ix}~u_{iy}]^T$
represents the Cartesian coordinates of the $i$th antenna element. Without loss of generality,
we assume antenna elements $i$ and $j$ lie on the y-axis and the impinging waveform spread
over $[\phi-\Delta,\phi+\Delta]$. Thus, for small $\Delta$ and antenna spacing $d_{ij}\defeq{u}_{iy}-{u}_{jy}$, we have
\begin{IEEEeqnarray}{rCl}
  [\mb{\Phi}]_{ij}
   &=&\int_{-\Delta}^{\Delta}p(\theta+\phi)e^{-j\frac{2\pi d_{ij}}{\lambda}\sin(\theta+\phi)}\mathrm{d}\theta\notag\\
   &\approx&
   e^{-j\frac{2\pi d_{ij}}{\lambda}\sin\phi}
   \int_{-\Delta}^{\Delta}p(\theta+\phi)e^{-j\frac{2\pi d_{ij}}{\lambda}\sin\theta\cos\phi}\mathrm{d}\theta. \notag
\end{IEEEeqnarray}
The integral on the right hand side is real if $p(\theta)$ is symmetric about $\phi$ and
for a system using ULA at the BS, $d_{ij}=(i-j)\xi$.

\setstretch{.97}


\begin{thebibliography}{15}
\bibitem{Scaling} F. Rusek, D. Persson, B. K. Lau, E. G. Larsson,
T. L. Marzetta, O. Edfors, and F. Tufvesson, ``Scaling up MIMO:
opportunities and challenges with very large arrays," \textit{IEEE
Signal Proces. Mag.}, vol. 30, no. 1, pp. 40--60, Jan. 2013.

\bibitem{Measurement2} S. Payami and F. Tufvesson, ``Measured
propagation characteristics for very-large MIMO at $2.6$ GHz," in
\textit{Proc. ACSSC}, Nov. 2012.

\bibitem{Lau} A. Liu and V. Lau, ``Joint power and antenna
selection optimization in large distributed MIMO networks," Tech.
Rep., 2012.

\bibitem{Kron_Model} J. P. Kermoal, L. Schumacher, K. I. Pedersen,
P. E. Mogensen, and F. Frederiksen, ``A stochastic MIMO radio
channel model with experimental validation," \textit{IEEE J. Sel.
Areas Commun.}, vol. 20, no. 6, pp. 1211--1226, Aug. 2002.

\bibitem{Hoydis} J. Hoydis, S. ten Brink, and M. Debbah, ``Massive
MIMO in the UL/DL of cellular networks: how many antennas do we
need?," {\it IEEE J. Sel. Areas Commun.}, vol. 31, no. 2, pp.
160--171, Feb. 2013.

\bibitem{SCM} ``Spatial channel model for multiple input multiple
output (MIMO) simulations," 3GPP TR 25.996 V11.0.0, Sep. 2012.
[Online]. Available:
http://www.3gpp.org/ftp/Specs/html-info/25996.htm

\bibitem{SCM_v} C.-X. Wang, X. Hong, H. Wu and W. Xu,
``Spatial-temporal correlation properties of the 3GPP spatial
channel model and the Kronecker MIMO channel model,"
\textit{EURASIP J. Wireless Commun. and Netw.}, 2007.

\bibitem{Yin} H. Yin, D. Gesbert, M. Filippou, and Y. Liu, ``A
coordinated approach to channel estimation in large-scale
multiple-antenna systems," {\it IEEE J. Sel. Areas Commun.},
vol. 31, no. 2, pp. 264--273, Feb. 2013.

\bibitem{Opt_pilot} M. Biguesh and A. B. Gershman, ``Training-based
MIMO channel estimation: a study of estimator tradeoffs and
optimal training signals,"  \textit{IEEE Trans. Signal Process.},
vol. 54, no.3, pp. 884--893, Mar. 2006.

\bibitem{Known_LSFG} L. Rong, X. Su, J. Zeng, Y. Kuang, and J. Li,
``Large scale MIMO transmission technology in the architecture of
cloud base-station," in \textit{Proc. IEEE GLOBECOM Workshops},
pp. 255--260, Dec. 2012.

\bibitem{Known_LSFG2} H. Q. Ngo, M. Matthaiou, and E. G. Larsson,
``Performance analysis of large scale MU-MIMO with optimal linear
receivers," \textit{Swedish Commun. Tech. Workshop (Swe-CTW)},
pp. 59--64, Oct. 2012.

\bibitem{Kay} S. M. Kay, \textit{Fundamentals of Statistical Signal
Processing: Estimation Theory}, Prentice Hall, 1993.

\bibitem{Pilot_con} J. Jose, A. Ashikhmin, T. Marzetta, and S. Vishwanath,
``Pilot contamination problem in multi-cell TDD systems," in {\it Proc. IEEE ISIT},
pp. 2184--2188, Jul. 2009.

\bibitem{LTE} ``Evolved universal terrestrial radio access
(E-UTRA); further advancements for E-UTRA physical layer aspects,"
3GPP TR 36.814 V9.0.0, Mar. 2010. [Online]. Available:
http://www.3gpp.org/ftp/Specs/html-info/36814.htm


\bibitem{Pilot_con2} F. Fernandes, A. Ashikhmin and T. L. Marzetta,
``Inter-Cell Interference in Noncooperative TDD Large Scale Antenna Systems,"
{\it IEEE J. Sel. Areas Commun.}, vol. 31, no. 2, pp.192--201, Feb. 2013.

\bibitem{RMT} R. Couillet and M. Debbah, \textit{Random Matrix
Methods for Wireless Communications}, New York, NY, USA: Cambridge
University Press, 2011.


\bibitem{Huh} H. Huh, G. Caire, H. C. Papadopoulos, and S. A.
Ramprashad, ``Achieving ``massive MIMO" spectral efficiency with a
not-so-large number of antennas," {\it IEEE Trans. Wireless
Commun.}, vol. 9, no. 5, pp. 3226--3238, Sep. 2012.

\bibitem{Bai} Z. D. Bai and J. W. Silverstein, \textit{Spectral Analysis of Large
Dimensional Random Matrices}, 2nd ed. Springer Series in
Statistics, New York, NY, USA, 2009.


\bibitem{Model_based} Y.-C. Chen and Y. T. Su, ``MIMO channel
estimation in correlated Fading Environments," \textit{IEEE Trans.
Commun.}, vol. 9, no. 3, pp. 1108--1119, Mar. 2010.

\bibitem{sim_model} D.-S. Shiu, G. J. Foschini, M. J. Gans, and J.
M. Kahn, ``Fading correlation and its effect on the capacity of
multielementantenna systems," \textit{IEEE Trans. Commun.}, vol.
48, no. 3, pp. 502--513, Mar. 2000.

\bibitem{Sayeed} A. M. Sayeed, ``Deconstructing multiantenna fading
channels," {\it IEEE Trans. Signal Process.}, vol. 50, no. 10,
pp.2563-2579, Oct. 2002.

\bibitem{Ozcelik}W. Weichselberger, M. Herdin, H. \"{O}zcelik, and
E. Bonek, ``A stochastic MIMO channel model with joint correlation
of both link ends," {\it IEEE Trans. Wireless Commun.}, vol. 5,
no. 1, pp. 90-100, Jan. 2006.


\bibitem{KLT} K. R. Rao and P. C. Yip, \textit{The Transform and
Data Compression Handbook}, CRC Press, Inc. Boca Raton, FL, USA, 2000.

\bibitem{KLT2} P. R. Haddad, A. N. Akansu, \textit{Multiresolution
Signal Decomposition, Second Edition: Transforms, Subbands, and
Wavelets}, Academic Press, Oct. 2000.

\bibitem{DCT} N. Ahmed, T. Natarajan, and K. R. Rao, ``Discrete
cosine transform," \textit{IEEE Trans. Comput.}, vol. C-23, no. 1, pp. 90--93,
1974.






\bibitem{Oppenheim} A. V. Oppenheim, R. W. Schafer,
\textit{Discrete-Time Signal Processing: Third Edition}, Pearson, 2010.






\end{thebibliography}
\end{document}